\documentclass{article}

\usepackage{multirow}%
\usepackage{amsmath,amssymb}%
\usepackage{amsthm}%
\usepackage{mathrsfs}%
\usepackage[title]{appendix}%
\usepackage{xcolor}%
\usepackage{textcomp}%
\usepackage{manyfoot}%
\usepackage{algorithm}%
\usepackage{algorithmicx}%
\usepackage{algpseudocode}%
\usepackage{listings}%

\usepackage{arxiv}

\usepackage[utf8]{inputenc} 
\usepackage[T1]{fontenc}    
\usepackage{hyperref}       
\usepackage{url}            
\usepackage{booktabs}       
\usepackage{amsfonts}       
\usepackage{nicefrac}       
\usepackage{microtype}      
\usepackage{lipsum}
\usepackage{graphicx}
\graphicspath{ {./images/} }

\title{Decentralized Identity Management on Ripple: A Conceptual Framework for High-Speed, Low-Cost Identity Transactions in Attestation-Based Attribute-Based Identity}

\author{
 Ruwanga Konara \\
  School of Computing \\
  University of Colombo\\
  Colombo, Sri Lanka \\
  \texttt{ruwangathandulakonara@gmail.com} \\
   \And
 Kasun De Zoysa \\
School of Computing \\
  University of Colombo\\
  Colombo, Sri Lanka \\
  \texttt{kasun@ucsc.cmb.ac.lk} \\
  \And
   Asanka Sayakkara \\
School of Computing \\
  University of Colombo\\
  Colombo, Sri Lanka \\
  \texttt{asa@ucsc.cmb.ac.lk} \\
}

\begin{document}
\maketitle
\begin{abstract}
The recent years have seen many industrial implementations and much scholastic research, i.e., prototypes and theoretical frameworks, in Decentralized Identity Management Systems (DIDMS). It is safe to say that Attestation-Based Attribute-Based Decentralized IDM (ABABDIDM) has not received anywhere near the same level of attention in the literature as general Attribute-Based DIDMs (ABDIDM), i.e, decentralized Attribute-Based Access Control (ABAC). The use of decentralization, i.e., DIDM, is to improve upon the security and privacy-related issues of centralized Identity Management Systems (IDM) and Attribute-Based IDMs (ABIDM). And blockchain is the framework used for decentralization in all these schemes. Many DIDMs - even ABDIDMs - have been defined on popular blockchains such as Hyperledger, Ethereum, and Bitcoin. However, despite the characteristics of Ripple that makes it appealing for an ABIDM, there is a lack of research to develop an Identity Management System (IDMS) on Ripple in literature. We have attempted to conceptualize an ABABDIDM on Ripple.
\end{abstract}


\section{Introduction}\label{sec1}

This study will progress with a surface-level review of  DIDM systems in early chapters and will shift its full focus to ABDIDM and decentralized ABAC halfway through the second chapter. Before the advent of blockchain, i.e., Distributed Ledger Technology (DLT), IDMs would be centralized solutions \cite{ahmed2022}. In a domain, a single authority or multiple authorities would act as central repositories of user data and authenticate users for users to be able to utilize resources from Service Providers (SP) in the domain, while providing other identity services such as identity creation, modification, verification, and revocation \cite{yuan2010}. However, there are privacy and security concerns about centralized IDMs, the root of which lie in this centralization and a trusted third-party being the identity manager. These issues include Single Point of Failure (SPOF), third-party exposure and monitoring (lack of privacy), where this third party is the identity service provider, and theft and abuse of private information of users by powerful online entities \cite{alanzi2022, seyam2023}. Hence, to mitigate these risks and compromises, blockchain-based IDMs were developed. There have also been scholastic research on blockchain-based ABDIDM and ABAC systems. There is research done one ABABDIDMS as well, although very rare. Centralized Attribute-Based IDM (ABIDM) also faces shortcomings such as SPOF, lack of user control since users do not own their attributes, exposure of user/ issuer/ verifier activity to the central host of the attribute repository, and monitoring, mutability in case of a breach, etc. Current work will be structured as following. The next section will cover the background: Decentralized IDM, Decentralized Attribute-Based IDM and ABAC, and Ripple and its suitability for identity management. Then the ABABDIDM on Ripple is described, after which come future directions for work. The last section concludes this work. 

\subsection{Revised Contribution in Light of New Decentralized Identifiers (DID) on Ripple}

In light of the recent XLS-40\cite{xrplxsl40} amendment on the XRP Ledger, which introduces native support for W3C DIDs, this work still makes contributions by advancing a layered identity-management design rather than merely defining identifiers. While XRPL’s DID method allows on-ledger storage or references of DID documents, it does not inherently address the full lifecycle of verifiable credentials, attribute attestations, revocation mechanisms, or privacy-preserving attribute disclosure. Our conceptual framework proposes an Attestation-Based, Attribute-Based Decentralized Identity Management (ABABDIDM) system that defines issuers, verifiers, and end-users in a high-throughput, low-cost architecture. We show how attribute attestations can be efficiently managed, how credential revocation can be supported, a sidechain for identity storage, a cross-chain interoperability approach, XRPL(and sidechain)-EVM bridging for logic execution, and how trust anchors might be governed, all grounded on XRPL’s performance and consensus structure. This design provides a path toward scalable and usable identity systems that go beyond basic DID resolution.

\section{Background}\label{sec2}

Traditional IDMs have three parties: Identity Provider (IDP), Service Provider (SP), and the user: \cite{torres2013}. The SP provides online services such as cloud, web applications, and any resources or facilities. The IDP authorizes and authenticates users so that they can access those services, i.e., provide identity services such as identity creation, modification etc. However, this study shall not explore DIDMs beyond the surface since our proposed system is not an IDM, but rather an attribute-based IDM.

\subsection{Decentralized Identity Management}

Blockchain was the framework that realized the design and implementation of DIDM. Most industrial implementations and academic prototypes of blockchain IDM have utilized a public or permissioned blockchain on Ethereum or Hyperledger: uPort \cite{lundkvist2016uport}, DNS-IDM \cite{kassem2019}, Health-ID \cite{javed2021}, \cite{cocco2021} are based on Ethereum, where \cite{modbus2021}, \cite{stamatellis2020}, and \cite{modbus2021} are on Hyperledger. There are examples such as Shocard \cite{shocard} that run on the Bitcoin ledger. According to \cite{dunphhy2018}, decentralized IDMs fall into two categories: Self Sovereign Identity (SSI) and Decentralized Trusted Identity (DTI). 

\subsubsection{Self Sovereign Identity}

Essentially, as self-governed collection of user information and identity attributes, these DIDMs eliminate the need for an IDP by giving users ownership and control of their identity \cite{seyam2023}. Individuals can determine what attributes (username, email, etc.) to be exposed, when to reveal, and to whom. Digital identities could be for persons, devices, institutions, etc. Sovrin \cite{sovrin2018whitepaper}, Health-ID \cite{javed2021},  uPort \cite{lundkvist2016uport}, and Microsoft ION are examples. 

\subsubsection{Distributed Trusted Identity}

It leverages pre-existing trusted credentials like government-issued IDs or passports. The proprietary service conducts identity proofing to authenticate these credentials and then records the verification proofs on the blockchain. This allows third parties to validate identities later. A notable example of this approach is ShoCard \cite{shocard}, which is based on Bitcoin's blockchain. This approach bears resemblance to ABABDIDM in the sense that both involve the issuance of attestations/proofs for user credentials/attributes by reputable parties for other parties to verify user credentials via these proofs.

\subsection{Attribute-Based Decentralized Identity Management}

Access to resources, i.e data or facilities, in ABIDM is not granted or denied by verifying nodes by maintaining Access Control Lists (ACL) \cite{ding2019}. It is done by assigning attributes to users. Therefore, user identity is a collection of attributes. There are three parties in an attribute-based IDM: user, attribute issuer, and verifier (service provider). A user applies for a particular attribute from an attribute issuer. The attribute issuer verifies the eligibility of the applicant to obtain this attribute, where the attribute is offered and recorded. But privacy is a concern, i.e, the verifying authority(SP), in some blockchain ABIDM schemes, shall not see the actual attribute but the proof of it.

Access can be given to the user if he satisfies the access requirement criteria of the SP, which is a set of attributes that the user should have in order for access/ authorization/ approval to be given. Decentralization is realized by the blockchain: no central server hosts, which avoids user tracking and SPOF. A group of trusted nodes can validate the ledger. In \cite{ding2019}, attribute issuers are the nodes responsible for maintaining the ledger through a vote-based consensus protocol. Their attribute-based access control system on blockchain for IoT is based on a predefined list of attributes, where these attributes and the transactions assigning them to users are stored on the blockchain and are visible to any entity. An SP can see and verify the actual attribute the user claims to have. It is interesting that all attribute-based decentralized IDMs in literature focus on access control. We were only able to find \cite{yang2022} as an ABDIDM that does not focus on access control but user-specific attributes such as income, ownership, etc. In \cite{YANG2024100199}, an Attribute-Based Access Control (ABAC) system based on Attribute-Based Encryption (ABE) running on Ethereum to protect IoT data is presented. This is a Ciphertext ABE (CP-ABE) scheme that uses the access policies of the data owner to encrypt the data, where the ciphertext and partially hidden access policy are stored on the cloud server and the blockchain, respectively. Attributes of the data user are hashed and stored on-chain. The decryption of the ciphertext is performed by secret keys generated using these attributes. All these functionalities are provided by smart contracts. Access revocation is not provided. A CP-ABE, ABAC system based on Ethereum is described in \cite{babu2023}, where they have added functionality for revocation, but data on the Inter-Planetary File Storage (IPFS) is not encrypted; rather, the access policy is encrypted, fully hiding the policy. Functionality is automated with a multitude of smart contracts. Unlike \cite{YANG2024100199}, this scheme does not rely on a gateway; rather, users directly interact with the blockchain, which is sensible since \cite{YANG2024100199} is for IoT devices. An ABAC system on Hyperledger Fabric with a Raft ordering is presented in \cite{shammar2022}, where validating nodes responsible for invoking smart contracts for policy operations, handling resource URLs on the blockchain, and processing access requests from the user are in several disjoint organizations. The data of data owners can be anywhere on the web, and the blockchain stores and manages the URL. There is a certificate authority (Hyperledger native) per organization to register any entity or device that would interact with the blockchain through the organization. Endorser nodes in organizations vote on new transactions; transactions invoke smart contract functions. Orderer nodes perform the consensus and block creation.

\subsubsection{Attestation-Based Attribute-Based Decentralized Identity Management}

In these schemes, the attribute is not stored on the chain, but an attestation (proof) verifying it from an issuer is. Privacy is a critical concern, i.e, the verifying authority, i.e, the SP, shall not see the actual attribute but the proof of it. A user applies for an attestation for a particular attribute from an attribute attestation issuer. The issuer verifies the eligibility, i.e, the attribute of the applicant, to grant this proof, where the attribute attestation is offered and recorded. As mentioned, the verifier, i.e, the SP, shall not see the actual attribute but the proof of it. For example, for a high-income individual, a bank could issue an attestation for the attribute income stating that this income is above a certain amount. A verifying bank could check that this attribute (income) is above a certain range by checking the attestation in order to approve a loan. Hence, attribute verification could be done without revealing the actual attribute which is private to the user \cite{yang2022}. In the ABDIDM system in \cite{yang2022}, since attributes are personal and private to a user, the attribute is not divulged to any party after the issuer signs the attestation for an attribute: the attribute is not stored on the chain, only the attestation. The verifier can use the attestation by verifying the issuer's signature on it, to check whether the user's statements about their attributes are true and if these attributes satisfy the approval criteria of the SP.

There are other attribute-based identity authentication systems that verify identity attributes without revealing additional information and authenticate users with anonymity \cite{mustafa2017, karl2019}. The motivation behind this exploration of distributed IDMs is to extract a framework basis for the ABABDIDM on Ripple.

\subsection{Ripple}

Ripple, in essence, is a decentralized currency exchange/remittance protocol. From a technical viewpoint, what differentiates Ripple from other blockchain protocols is its low-cost, vote-based consensus algorithm that involves a trusted set of validating nodes known as servers \cite{schwartz2018ripple}. To further accelerate the consensus process, each server only considers and votes on proposals (transactions to be validated) from a subset of servers known as the Unique Node List (UNL) of the server. This energy-efficient and rapid consensus algorithm runs every few seconds, enabling thousands of transactions per second \cite{jani2018}. The XRP ledger is the distributed ledger maintained by this consensus protocol, which records transactions and holds account balances: XRP balances, fiat currency balances in the form of IOUs \cite{rosner2016ripple}. The cross-currency, i.e., cross-border, transaction model requires two actors: market makers such as hedge funds and currency exchange desks who provide liquidity between assets and financial institutions (banks) that hold funds and issue balances or payment processors that issue balances on behalf of customers. Market makers can be hedge funds, currency exchanges, or financial institutions that provide liquidity between different currencies. They facilitate on-demand liquidity (ODL) by exchanging fiat currencies into XRP (and vice versa) to bridge transactions efficiently. Banks or payment processors hold accounts and issue balances for customers. For cross-border transactions, they either use RippleNet’s messaging system, i.e., sending funds as IOUs, to settle transactions or leverage ODL for real-time settlements using XRP. RippleNet is the network of Ripple nodes. XRP can be sent directly between individuals' XRPL accounts without intermediaries. 

While technically each server can choose its own UNL, most adopt Ripple’s recommended UNL for safety and consistency. It is worth noting that this is a practical centralization trade-off. RippleNet \(\neq\) XRPL; RippleNet is a private network managed by Ripple Inc., whereas XRPL is a public ledger.

There is an individual who wants to send USD such that the receiver can get EUR. They have banks that are partners with Ripple. For a cross-border fiat transaction, there is a few approaches. 

\begin{itemize}
\item Using XRP as a Bridge (ODL): If the bank uses ODL, the process works as follows. The sender (from America) initiates a USD transfer, i.e., their bank converts USD to XRP. XRP is sent instantly across the XRP Ledger (XRPL). The receiving bank (Europe) converts XRP into EUR. The recipient gets EUR in their bank account.

\item Using RippleNet’s Messaging System (No XRP Involved): If banks don’t use XRP but are part of RippleNet, the transaction works as follows. Sender’s bank (USD) sends a payment request via RippleNet. RippleNet finds a direct liquidity provider (another bank or market maker) to exchange USD to EUR. EUR is sent to the recipient’s bank in Europe. Recipient receives EUR in their account.

\item Using Gateways (Ripple’s Built-in Exchange Mechanism on XRPL): Another method is using gateways, which are entities that issue IOUs (representing fiat currency) on the XRP Ledger (XRPL). The sender deposits USD with a Ripple gateway (e.g., Bitstamp, GateHub). Gateway issues USD IOUs (a digital representation of USD) on the XRPL. The recipient redeems the IOUs for EUR through another gateway in Europe. The recipient gets EUR in their bank account.

\end{itemize}

A local USD to USD to transaction:

\begin{itemize}
    \item  On-Chain (XRP Transfer, Bank Not Required): A sender transfers XRP to a recipient’s XRP wallet. The recipient can hold XRP or convert it to USD through an exchange. Banks are not directly involved in this process.

    \item Off-Chain (RippleNet, Bank-Driven): If both sender and recipient have accounts at RippleNet-enabled banks, the banks handle the transaction. The sender’s bank debits USD and communicates via RippleNet to the recipient’s bank. The recipient’s bank credits the USD instantly without waiting for traditional settlement. XRP is not involved, but this is a faster, cheaper alternative to SWIFT.
\end{itemize}

Ripple has certain features that enable fast and low-cost, cross-border transactions.

\begin{itemize}
    \item Pathfinding – Finding the Best Route: When someone wants to send one currency and receive another, the XRPL automatically finds the best (cheapest) path by checking direct liquidity between the two currencies (e.g., USD → EUR), liquidity through intermediate assets (e.g., USD → XRP → EUR), liquidity across multiple order books (multi-hop transactions), i.e., find the cheapest path across multiple market makers over a chain of different assets: \cite{rosner2016ripple, jani2018}. Alice wants to send USD to Bob, who wants EUR. The XRPL looks at all available liquidity paths (USD → EUR, USD → XRP → EUR, etc.). It picks the cheapest and fastest route automatically. Pathfinding = Best rate, lowest fees, and fastest transaction. This means that in cross-border payments, a market maker in Ripple's exchange is a foreign-exchange trader who facilitates currency trades by posting bids and offers. They function like market makers in other markets, profiting from the spread between buy and sell prices. The Ripple protocol optimizes transactions by selecting both the cheapest offer and the most efficient path, driving market makers to compete for better spreads while ensuring liquidity \cite{rosner2016ripple}.

    \item Auto-Bridging – XRP as the Middleman; If there is no direct liquidity between two assets, XRP can act as a bridge. This increases liquidity for assets that do not have enough direct orders. It lowers costs by using XRP’s high liquidity and low fees. Imagine there is no direct market for USD/EUR. But there is USD/XRP and XRP/EUR liquidity. The XRPL auto-bridges the trade: USD → XRP → EUR seamlessly. Auto-Bridging makes transactions cheaper and faster by using XRP liquidity.

    \item XRPL’s Decentralized Exchange (DEX): The XRPL has a built-in DEX where people can trade assets (USD, EUR, BTC, etc.), provide liquidity (market makers place buy/sell orders), and use automated order matching. There are market makers and order books; market makers place bids (buy orders) and asks (sell orders), the DEX matches the best orders across different paths, and anyone can be a market maker by placing liquidity in the order books.

\end{itemize}

\subsubsection{Ripple's suitability for Attestation-Based Attribute-Based Identity Management}

Popular blockchains such as Ethereum, Hyperledger, and Bitcoin have all been employed as frameworks for decentralized identity management as cited above, but not necessarily attribute-based identity management. However, despite XRP being consistently among the 5 biggest cryptocurrencies on the market \cite{coinmarketcap}, there is no published academic or industry research on any sort of identity management on Ripple; we have not been able to find any such work. Despite not having been explored yet, Ripple has certain features that make it appear ideal for ABABDIDIM.

Suitability for Attestation-Based Attribute-Based DIDM

\begin{itemize}
    \item Selective Disclosure and Privacy: Identity attributes (age, nationality, income, etc.) can be stored off-chain while their proofs are hashed and stored on XRPL. Users can selectively share only the necessary attributes (e.g., proving they are over 18 without revealing their birthdate).

    \item  Smart Contracts (Hooks) for Automated Access Control: XRPL’s Hooks (lightweight smart contracts) allow automated validation of identity attributes. For example, a nightclub could have a Hook that only allows users with a "21+ Token" to make a purchase. In addition, "Ripple is enhancing the XRP Ledger by integrating Ethereum-compatible smart contracts via a new sidechain, expanding its functionality beyond basic transactions to include complex applications like decentralized exchanges and token issuance", according to CoinDesk \cite{coindesk2024ripple}. By April 2025, Ripple had already integrated Ethereum Virtual Machine (EVM)-compatible smart contracts via a sidechain
    
    \item Multi-Signature Identity Verification: XRPL supports multi-signature accounts, meaning multiple entities (banks, government agencies) can co-sign to verify a person’s attributes. For example, a job applicant may need signatures from both their university and their employer before getting a work visa.

    \item Attribute-Based Access Control (ABAC) with Trust Lines: XRPL’s trust line system allows organizations to issue and verify attribute-based credentials. For example, a university can issue a "Student ID Token", and a discount provider verifies it before granting student benefits.

    \item  Ripple’s Banking Partnerships: For ABIDM, Ripple’s deep ties with banks and regulated financial networks ensure that attribute-based identity verification can be trusted and recognized internationally. Banks already categorize customers based on account type, credit score, transaction history, and regulatory status, which can be tokenized as identity attributes on XRPL. Banks can issue attribute-based credentials such as \textbf{\textit{Verified Business Owner}}, \textbf{\textit{High Net Worth Individual}}, and \textbf{\textit{Taxpayer}} that can be used across financial services. Cross-border financial compliance can be automated using ABIDM tokens instead of manual verification. Institutions can share verifiable attributes on XRPL without exposing sensitive personal data. For example, a business owner verified by a Ripple-partnered bank can receive a \textbf{\textit{Certified Business Owner}} token on XRPL. This could be used to access special business loans, premium banking services, or international trade benefits without submitting paperwork repeatedly.

\end{itemize}

\section{Attestation-Based Attribute-Based IDM System on Ripple}

There is a multitude of blockchain-based ABACs in the literature and industry. These are defined on versatile, flexible blockchains with extensive smart contract functionality or blockchain frameworks specifically crafted for IDMs. They provide secure, authorized access to data or services. Ripple, on the other hand, is a highly domain/task-specific and rigid framework. Therefore, it is not sensible to develop an access control protocol on Ripple. However, with Ripple's ties to the financial infrastructure, it may be ideal for an attestation and credential-based identity framework. However, recently, Ripple has integrated Ethereum-compatible smart contracts into the XRP ledger via a side-chain \cite{Mitrade2025}, opening possibilities for decentralized apps to be developed on Ripple. These smart contracts that would execute on the EVM would be instrumental in adding functionality and autonomy to a possible IDM on Ripple. 

Ripple has established partnerships with various government institutions, primarily focusing on the development of Central Bank Digital Currencies (CBDCs). These include government financial authorities in Bhutan \cite{ripple_bhutan2021}, Palau \cite{ripple_palau2023}, Georgia \cite{ripple_georgia2023}, \cite{ripple_montenegro2023}, and discussions with other governments and government agencies \cite{ripple_governments2023}. This opens possibilities for other government agencies (non-financial) to collaborate with Ripple for a potential IDM system. 

\subsection{IDM model}

There are three roles to be identified: the user to whom an identity with attributes belongs, the authority (bank, government organization) that issues attestations for users' statements, i.e., claims, on their attributes, and the verifier who needs to verify certain attributes of the user in order to provide a service to them. 

The idea is to have users control their own identities and use them without revealing information that is unnecessary to be disclosed (self-sovereign). The scheme runs on an identity management sidechain with EVM compatibility. This is a modified version of the XRPL EVM, i.e., the Ripple sidechain compatible with the EVM. On the sidechain, as necessary in most cases involving sidechains, an account for the user is created. This account holds the same information as the XRPL account of the user, including public key and the address on the XRPL account, balances such as \textbf{IdXRP} tokens, XRP balances, etc. In addition, a user account contains addresses of its attestations on the blockchain. This may be viewed as similar to the XRP or any asset balance on a Ripple account. These addresses may be viewed as a type of balance from attestation issuance transactions received by the user. The network would be running the Proof of Authority (PoA) consensus, i.e, CometBFT (Tendermint), that dictates the ledger of the original EVM-compatible sidechain on Ripple. For cross-chain communication and asset (XRP) transfer between this sidechain, the Axelar network is used. The XRPL EVM bridge is used for asset movement between the XRP Ledger and the EVM sidechain. Similar to the original XRPL EVM sidechain, this sidechain is also built on the Cosmos SDK.

\#\#\#\# - \textit{This \textbf{IdXRP} \ref{econ} is a token native to the IDM sidechain.}

 \begin{figure}[ht]
\centering
\includegraphics[width=0.9\textwidth]{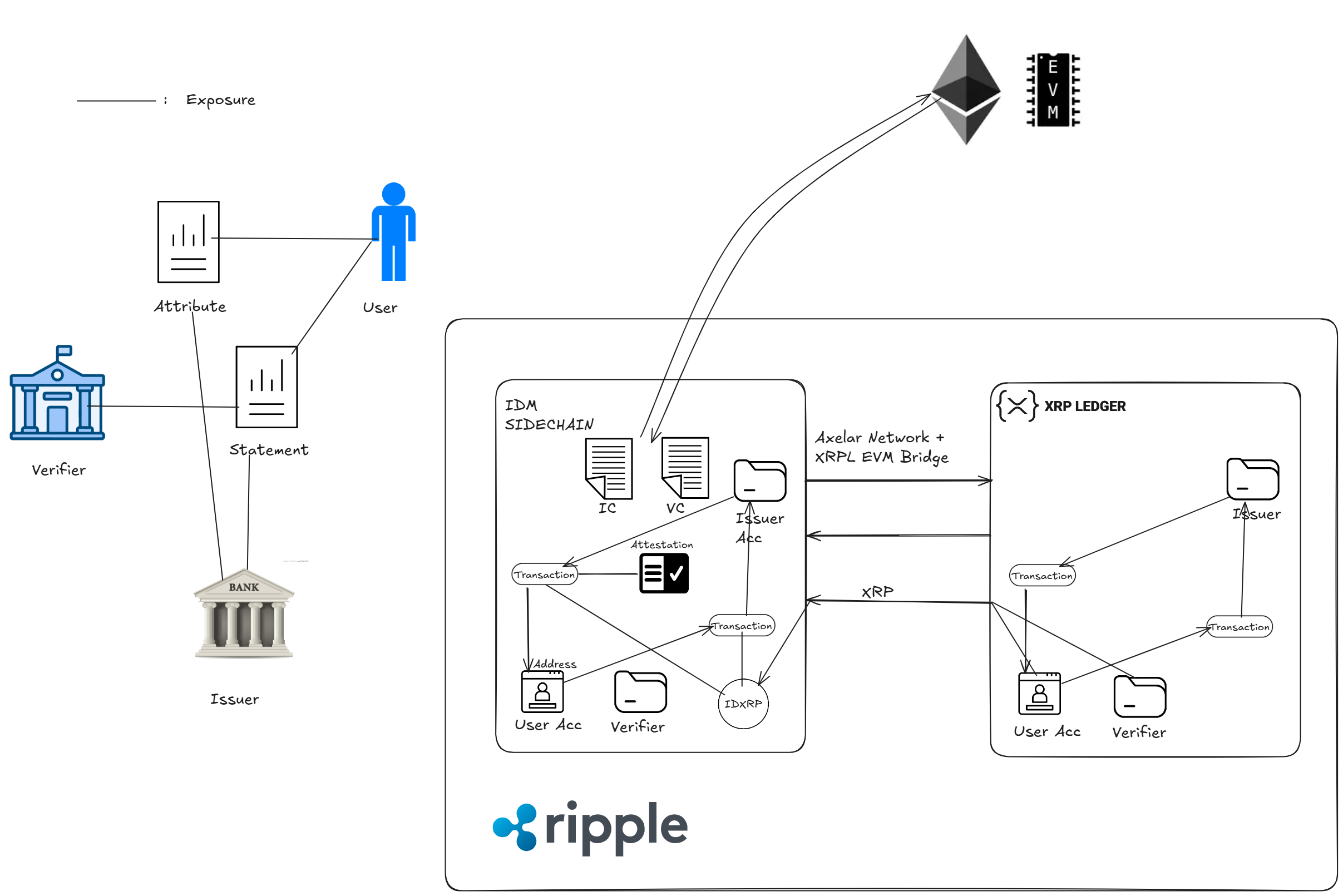}
\caption{System Model}\label{model}
\end{figure}

\subsubsection{Smart Contracts}

\begin{figure}[ht]
\centering
\includegraphics[width=0.9\textwidth]{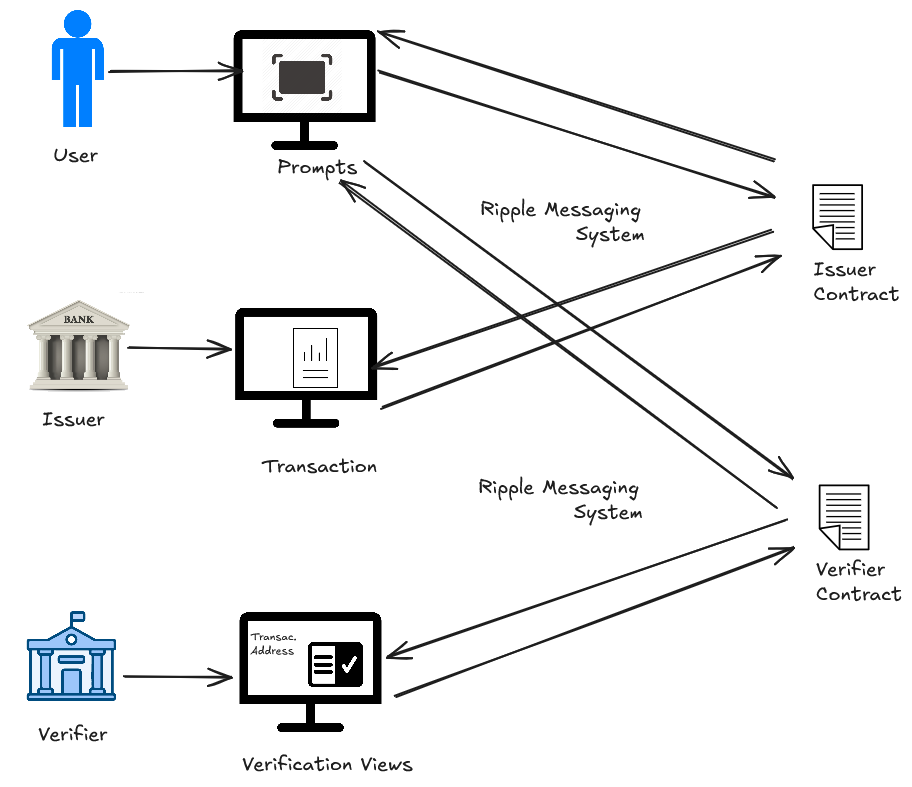}
\caption{Interactions of Smart Contracts and Entities}\label{interactions}
\end{figure}

There are 2 smart contracts in the side chain as shown in \textbf{Figure \ref{model}}: Issuer Contract (IC) and Verifier Contract (VC). IC is responsible for issuing attribute attestations and is triggered by transactions from attestation issuers, whereas VC is responsible for verifying the attributes upon execution of transactions from verifiers. These contracts execute on the EVM. To comply with Ripple's transaction model, each attestation issuance operation transaction is coupled with a \(0.3XRP \) transfer on the XRP ledger, executed automatically through the smart contract to validate the transaction and ensure finality on XRPL, to prevent transaction flooding, DoS, etc. \textbf{Figure \ref{interactions}} shows interactions between entities and smart contracts.

\subsubsection{Example Scenarios}

Users must be able to prove that they satisfy the conditions set by identity verifiers without revealing their actual identity information. The identity attributes must be private to the owners. Users keep their identity attributes private and use only the on-chain attestations to prove statements about their identity attributes to verifiers.

\begin{enumerate}
    \item Scenario 1: There is a Ministry of Agriculture, and they only license entities that own and farm more than 20 hectares of wheat to apply a certain mineral fertilizer X. However, it is unnecessary for the ministry to be disclosed the actual area of wheat the entity Rex owns. This attribute (the amount of wheat fields owned by Rex) is provided an attestation for by the Department of Federal Income. The statement of the user is that they own above 20; they have 31 hectares. The attestation is issued and signed by the Dept. of Fed. Income, whereas the Ministry of Agriculture is the verifier and SP. The attestation should bear that they farm above 20, not that they farm 31. The ministry would be disclosed only the necessary information, which is that they own more than 20 hectares.

    \item Scenario 2: A certain credit facility is offered by Bank of A to new or old customers with at least  \(70,000\$\) liquidity at a reputable institution. This "Reputation" of the issuer is an important attribute for a verifier in this niche of ABIDMs. Since Ripple's partners, current and potential, are reputed financial institutions and government authorities, metrics in our IDM for the reputation we deem unnecessary. A customer with more than \(40,000 \$\) at Bank of B and more than \(55,000 \$\) at Bank of C has statements "More than \(25,000 \$\) at B" and "More than  \(45,000 \$\) at C." The respective banks issue attestations for these claims. Verifier A can see that the customer exceeds the liquidity threshold without discovering their actual amount of liquid funds at other institutions; the customer has no funds at A since he is a new customer. 
\end{enumerate}

\subsubsection{Scenario 1} \label{one}

 The Department of Federal Income has validated the statement from Rex declaring that they farm more than 20 hectares. They know that he has 31 hectares of wheat. They initiate a transaction on the sidechain with the usual Ripple transaction syntax. There needs to be a field for the signed statement. They include the statement when initiating the transaction. The IC signs the statement with the secret key of the Federal Income, i.e., the key that signs the transaction, and, using the Ripple messaging system, prompts Rex to confirm or deny the validity of the statement. Once he validates the statement, IC prompts Rex to sign the statement, resulting in a double-signed statement. Then the transaction is submitted by the IC. On the main network, for the XRPL, a corresponding transaction of \(0.3XRP\) is submitted by the IC. These transactions are from Federal Income to Rex. On the sidechain, the IC stores the address of the sidechain transaction on Rex's account. And the IC then submits the reimbursement transaction of \(0.3XRP\) from Rex to Federal Income on the XRPL and also on the sidechain to reimburse IdXRP costs; the model of economy \ref{econ} describes related details including asset transfers between the two chains. For these transactions on the IDM sidechain, XRP is moved to the IDM sidechain and wrapped as IdXRP tokens: this movement takes place through the XRPL EVM bridge and the Axelar network. These signatures are Elliptic Curve Digital Signature (ECDSA) secp256k1 signatures, which is the signature scheme on Ripple, and the account on the sidechain provides the keypair out of which the secret key is used to sign. Through the verification views of the frontend application (\textbf{Figure \ref{interactions}}), the ministry submits the address it received from Rex. The VC then loads the details of the transaction, the statement, the signature verification results, and the identities of Federal Income and Rex. Although the XRP ledger is pseudo-anonymous in operations, this IDM sidechain openly shares identities because the reputation of the issuer determines the credibility of the attestation. The attestation is the double-signed statement (See \textbf{Figure \ref{model}}). This verification process involves no transactions and, also the VC triggers no transactions on either chain; however, there may be IdXRP transactions invoked by the VC as described in \ref{econ}. The ministry can see that Rex has more than 20 hectares of wheat, hence the ministry can issue the license for the purchase of X. The ministry does not discover the actual attribute that belongs to Rex (\textbf{Figure \ref{model}}), which is that he owns 31 hectares of wheat. This is not even stored on chain; only the attestation regarding the user statement regarding the attribute is.

 \subsubsection{Scenario 2}

 The banks of B and C have the statements given previously. Both institutions know that these statements are valid. Identical to the above scenario, B initiates the transaction and signs the statement, followed by the customer's signature, and corresponding transactions are executed with \(0.3XRP\) as the transaction volume. The same takes place regarding the attestation from C. The addresses of both these attestation issuance transactions are stored on the customer's account by the IC, which he shares with the Bank of A. The VC would pull and display these transaction details: statements; attestations; signature verifications; and the identities of B, C, and the customer. By verifying the legitimacy of the attestation, i.e., the double signature verifications, it is confirmed by A that the customer has liquidity more than \(70,000\$\). Hence, the credit facility can be approved for the customer.



 \subsection{Model of Economy} \label{econ}

The token native to the IDM sidechain, \textit{IdXRP}, is a wrapped representation of XRP used both for transaction value and gas fees within the sidechain (\textbf{Figure \ref{model}}). When the issuer submits an attestation issuance transaction, it is processed by the IC as described. A corresponding transaction of \(0.3\) XRP is made on the XRPL from the issuer to the user.

To mirror this on the IDM sidechain, \(0.3\) XRP is moved to the sidechain, wrapped as IdXRP, and included in the issuance transaction. Since the IDM sidechain operates an EVM-compatible environment, smart contracts consume gas analogous to Ethereum's gas model. However, the gas fees are paid in IdXRP, calculated based on the equivalent ETH value consumed during IC execution. This expenditure is borne by the issuer, including the cost for reimbursement phase IC execution; this cost too is reimbursed as described below.

During the reimbursement phase, a \(0.3\) XRP transaction occurs from the user to the issuer on the XRP Ledger. For the corresponding transaction on the IDM sidechain, \(0.3\) XRP plus the XRP-equivalent of the IC's gas usage is moved from the user's XRP Ledger account, wrapped as IdXRP, and sent to the issuer. This dual-ledger setup ensures auditability and maintains consistency between the XRPL and the IDM chain.

\#\#\#\# - \textbf{If the verifier does not have sufficient IdXRP for the IC gas cost, his XRP on the sidechain would be wrapped as IdXRP. If there is insufficient XRP for this purpose, XRP would be bridged to the issuer's sidechain account from the XRPL. The same applies for the \(0.3\) XRP that needs to be wrapped as IdXRP to be sent to the user. In the reimbursement phase, only bridge and/or convert assets if the user's account on the sidechain lacks sufficient amounts of respective funds. All conversions and bridging that may happen on the issuer's side due to the insufficiency of the respective funds shall be reversed upon the completion of the reimbursement phase.}

\#\#\#\# - If the user verifies the statement as invalid, two transactions from the issuer to the user take place: one on the XRPL carrying \(0.3\) XRP and another on the sidechain carrying \(0.3\) XRP wrapped as IdXRP. The gas IdXRP is spent by by the issuer. This penalization aids in maintaining the IDM chain functioning without Denial of Service \ref{dos}


During the attestation verification process, the verifier initiates the verification by triggering the VC. This invocation does not immediately proceed to computation. Instead, a prompt is issued to the user associated with the attestation, requesting confirmation of the verifier's access to the attested statement.

Verification cannot proceed without the user's explicit approval. This ensures that control over identity data remains decentralized and user-centric.

The economic model of verification incorporates an incentive alignment mechanism:

\begin{itemize}
  \item If the user \textbf{denies} the request, terminating the verification process, the verifier bears the gas cost, and an equivalent amount of IdXRP is exacted from their balance to cover the invocation of the VC. If the verifier does not have sufficient IdXRP, his XRP on the IDM chain would be wrapped into IdXRP. If the sidechain does not have sufficient XRP, XRP would be bridged from the XRPL account.
  
  \item If the user \textbf{approves} the request and verification is executed successfully as described in \ref{one}, the user assumes the gas cost from the point of the user's approval onward. The cost borne by the verifier is reimbursed by the VC via an IdXRP transaction from the user to the verifier on the sidechain. Similar to the previous case, initially, the cost is borne by the verifier, where conversions and bridging would happen where necessary. Upon reimbursement, all such conversions would be reversed as well.
\end{itemize}

This dual-mode model ensures that:
\begin{itemize}
  \item Verifiers are penalized for unnecessary or spam-like verification attempts.
  \item Users are incentivized to participate in legitimate verifications.
  \item The computational burden aligns with approval, thus reinforcing consent-driven data exchange.
\end{itemize}

\#\#\#\# - \textbf{This model of economy centered around the \(0.3\) XRP and \(0.3\) XRP in IdXRP from the issuer discourages spam flooding of transactions by the issuer. In the verification scenario, the verifier is discouraged from such behavior as described above. Therefore, the IDM model does not have the need for transaction fees that ripple transactions incur in the traditional sense, i.e., this IDM does not have transaction fees beyond the asset transfers described. However, there may be costs related to cross-chain assets bridging in some cases.} \label{dos}

\section{Future Directions}

The sidechain used is a modified version of the XRPL EVM sidechain. Considering Ripple's suitability for an ABABDIDM of the given genre, the Ripple community may consider a prototype. Another approach may be to consider Ripple's ties with existing financial infrastructure in devising an ABAC on a Ripple sidechain for financial databases. Given Ripple’s integration with existing financial infrastructure, its fast settlement capabilities, and regulatory-oriented design, a range of impactful systems could be developed. For example, a system may perform identity-based instant loan approvals, where verified identities on an IDM chain enable rapid underwritten credit issuance tied directly to financial institutions already connected to RippleNet. Cross-border micropayments for verified users could be streamlined, enabling remittances and Business to Business (B2B) payments to occur within seconds while complying with regulatory requirements. Compliance and audit reporting tools for financial institutions could also be developed on Ripple sidechains, allowing real-time reporting of verified transactions to regulators. Additionally, tokenized representations of traditional financial instruments, such as bonds, insurance products, or trade finance assets, could be launched with identity-backed assurance, enhancing trust and liquidity. Overall, Ripple’s positioning as a bridge between blockchain and traditional finance offers unique opportunities to scale systems that require high trust, low cost, and speed.

\subsection{Limitations}

While the proposed Attestation-Based, Attribute-Based Decentralized Identity Management (ABABDIDM) framework builds on XRPL’s XLS-40 DID capabilities and outlines a conceptual pathway for cross-chain identity management, several limitations must be acknowledged:

\begin{enumerate}
    \item \textbf{Formal Security Analysis Not Included.} This work is conceptual and does not provide formal proofs for security against adversarial scenarios, such as colluding issuers, malicious verifiers, or sidechain compromise.
    
    \item \textbf{Privacy Guarantees are Conceptual.} Although the framework considers privacy-preserving attribute disclosure, detailed protocols (e.g., zero-knowledge proofs) are not implemented, and full unlinkability is not formally established.
    
    \item \textbf{Operational Architecture is Abstract.} The IDM sidechain, cross-chain interoperability, and XRPL- and sidechain-EVM bridging are described at a high level; deployment strategies, transaction throughput, and cost evaluations remain future work.
    
    \item \textbf{Trust and Governance.} While the framework outlines trust anchors and governance mechanisms, detailed incentive structures and policies for issuers, verifiers, and end-users are not fully developed.
    
    \item \textbf{Revocation and Lifecycle Management.} The approach to credential revocation and lifecycle management is proposed conceptually; efficient on-chain/off-chain revocation mechanisms require further research.
    
    \item \textbf{Dependency on XRPL Infrastructure.} The framework assumes XLS-40 DID infrastructure and XRPL ledger stability; future changes to XRPL consensus or DID standards may require adaptation of the proposed system.
    
    \item \textbf{Preliminary Conceptual Stage.} The work is meant to communicate the architecture and design principles for a high-throughput, low-cost identity system. Full implementation, testing, and benchmarking are deferred to future research.
\end{enumerate}

\section{Conclusion}

We have discussed the necessity of decentralized identity management and its realization with the emergence of the DLT. An acknowledgment of DIDMs with its subclasses has been followed by a survey of ABDIDMs, where a collection of ABAC systems and Attestation-Based ABDIDM systems have been explored. A detailed description of Ripple followed, along with its suitability for an ABDIDM, setting the landscape for an attestation-based ABDIDM on Ripple. The proposed system was then described.
By leveraging Ripple’s fast settlement, regulatory alignment, smart contract integrations, and integration with traditional finance, the proposed IDM platform represents a practical and scalable solution for decentralized attestation and attribute-based identity. Future work may involve implementing a prototype, further optimizing the gas economy, and exploring extensions into areas such as compliance automation, tokenized financial instruments, and real-time audit trails, all rooted in Ripple’s enterprise-ready infrastructure.

\bibliographystyle{unsrt}  






\end{document}